\documentstyle[12pt,aaspp4]{article}

\begin{document}

\title{CAN FIREBALL OR FIRECONE MODELS EXPLAIN GAMMA RAY BURSTS?} 
\author{Arnon Dar}
\affil{ Max Planck Institut F\"ur Physik\\ 
(Werner Heisenberg Institut)\\
Fohringer Ring 6, 80805 M\"unchen, Germany\\ 
and\\ 
Department of Physics and
Space Research Institute\\ 
Israel Institute of Technology, Haifa 32000,Israel\\ 
arnon@physics.technion.ac.il } 

\begin{abstract} 
The observed afterglows of gamma ray bursts (GRBs), in particular that of
GRB 970228 six months later, seem to rule out relativistic fireballs and
relativistic firecones driven by merger or accretion induced collapse of
compact stellar objects in galaxies as the origin of GRBs. GRBs can be
produced by superluminal jets from such events. 
\end{abstract} 

\section{INTRODUCTION} The isotropy of the positions of gamma ray bursts
(GRBs) in the sky and their brightness distribution have provided the
first strong indication that they are at cosmological distances (Meegan et
al 1992; Fishman and Meegan 1995 and references therein). The recent
discovery of an extended faint optical source coincident with the optical
transient of GRB 970228 ( Groot et al. 1997; van Paradijs et al. 1997;
Sahu et al. 1997) and, in particular, the detection of absorption and
emission line systems at redshift z=0.835 (Metzger et al. 1997a,b) in the
spectrum of the optical counterpart of GRB 970508, which may arise from a
host galaxy (see e.g. Pedersen et al 1997), have provided further evidence
that GRBs take place in distant galaxies.  The peak luminosity of GRB
970508 in the 0.04-2.0 MeV range exceeded $10^{51}d\Omega~erg~s^{-1}$
(assuming $\Omega\approx 0.2$, $\Lambda=0$ and $H_0\approx
70~km~Mpc~s^{-1}$), where $d\Omega$ is the solid angle into which the
emitted radiation was beamed.  Such $\gamma$-ray luminosities and their
short time variability strongly suggest that GRBs are produced by mergers
and/or accretion induced collapse (AIC) of compact stellar objects
(Blinnikov et al. 1984; Paczynski 1986; Goodman, Dar and Nussinov 1987),
the only known sources which can release such enormous energies in a very
short time. Then the gamma rays must be highly collimated and their radius
of emission must be large enough in order to avoid self opaqueness due to
$\gamma \gamma\rightarrow e^+e^-$ pair production. A sufficient, and
probably necessary, condition for this to occur is that they are emitted
by highly relativistic outflows with bulk Lorentz factors,
$\Gamma=1/\sqrt{1-\beta^2}\gg 100.$ Additional support for their emission
from highly relativistic flows is provided by their non thermal energy
spectrum.  Consequently relativistic fireballs (Cavallo and Rees 1976;
Paczynski 1986; Goodman 1986) and relativistic jets (e.g., Shaviv and Dar
1995; Dar 1997) were proposed as the producers of GRBs. The observed
radiation may be produced by self interactions within the flow (e.g.,
Paczynski and Xu 1994; Rees and Meszaros 1994) or by interactions with
external matter (e.g. Rees and Meszaros 1992; Meszaros and Rees 1993) or
with external radiation (e.g., Shemi 1993;  Shaviv and Dar 1995; 1996). 

Following the discovery of the afterglow of GRBs 970228  
various authors have concluded that it supports the fireball model of GRBs
(e.g., Katz et al. 1997; Waxman 1997a,b; Wijers et al. 1997; Reichart
1997; Vietri 1997; Rhoads 1997; Sari 1997a; Tavani 1997; Sahu et al
1997a). However, here we show that the detailed observations of the
afterglows of GRBs 970111, 970228, 970402, 970508, 970616, 970828, 971214,
and in particular that of 970228 six months later (Fruchter et al,
1997), support neither the simple relativistic fireball model (e.g.,
Meszaros and Rees 1997), nor simple relativistic firecone (conical ejecta)
models. However, if the relativistic ejecta in merger/AIC of compact
stellar objects is collimated into magnetically confined narrow jets, the
major problems of the fireball and firecones models can be avoided and the
general properties of GRBs and their afterglows can be explained quite
naturally. 

\section{FAILURES OF SIMPLE FIREBALLS}
\subsection{Energy Crisis} 
The spherical blast wave models assume (e.g., Meszaros and Rees 1997; 
Wijers et al 1997) that the ultrarelativistic spherical shell which
expands with a Lorentz factor $\gamma=1/\sqrt{1-\beta^2}$ drives a
collisionless (magnetic) shock into the surrounding interstellar medium. 
They also assume that the collisionless shock which propagates in the ISM
with a Lorentz factor $\gamma_s=\sqrt{2}\gamma$ accelerates it and heats
it up to a temperature $T\approx \gamma m_pc^2$ (in its rest frame). 
Energy-momentum conservation in the ultrarelativistic limit, which reads
$d[(M+nm_p(4\pi/3)r^3)\gamma^i]\approx 0$ with $i=2$, then implies that
the bulk Lorentz factor of the decelerating debris (mass $M$) and swept up
ISM (ambient density $n$)  decreases for large $r$ like
$\gamma(r)\sim\gamma(r_0)(r/r_0)^{-3/2}$.  In fact, the assumption that a
highly relativistic collisionless shock heats up the ISM to a temperature
$T_p\approx \gamma m_pc^2$ in its rest frame, has never been substantiated
by self consistent magnetodynamic calculations nor by direct observations
of radiation from decelerating superluminal jets.  For $T_p<m_pc^2$ (or
fast cooling) one has $i=1$ and $\gamma \sim r^{-3}$. It is further
assumed that superthermal electrons, with a power-law spectrum,
$dn_e/dE'\sim E'^{-p}$ and $p\approx 2.5$, in the rest frame of the
shocked ISM emit synchrotron radiation with a power-law spectrum $h\nu
dn/d\nu'\sim
\nu'^{-(p-1)/2}$ (or $\sim \nu'^{-p/2}$ for fast cooling) from an assumed
equipartition internal magnetic field. 
Photons which are emitted with a frequency $\nu'$ in the rest frame of the
shocked material and at an angle $cos\theta'$ relative to its bulk motion,
are viewed in the lab frame at a frequency $\nu$ and at an angle $\theta$
which satisfy, respectively, (e.g. , Rybicki and Lightman 1979) 
\begin{equation} 
\nu=\gamma(1+\beta cos\theta')\nu'~;~~
tan\theta=sin\theta'/\gamma(\beta+cos\theta'). 
\end{equation} 
If $\gamma>>1$ and if the photons are emitted isotropically in the rest
frame of the shocked material with differential intensity
$I_\nu= h\nu'dn/d\nu'$, then
in the lab frame they have an angular and spectral distribution
\begin{equation}
{dI_\nu\over dcos\theta}= {4\gamma^3\over(1+\gamma^2\theta^2)^3}
I_{\nu'=\nu(1+\gamma^2\theta^2)/2\gamma}. 
\end{equation} 
Thus, a distant observer sees essentially only photons
emitted in his direction from radius vectors $r$ with angles $\theta\leq
1/\gamma$ relative to his line of sight (l.o.s.) to the explosion center
($r=0)$. If the emission from the shocked ISM between the expanding debris
and the shock front is uniform, then the photon arrival times are
\begin{equation} 
t\approx {r\over \alpha_ic[\gamma(r)]^2}-{r'\over
2\alpha_ic[\gamma(r')]^2}+ {r\theta^2\over 2c}, 
\end{equation} 
where $r'\leq r$ is the initial distance of the shocked material from the
explosion point and $\alpha_i=2(6/i+1)=14,8$ for $i=1,2,$ respectively. If
the photons are emitted mainly from a thin shell behind the shock front
then $r'\approx r$ and 
\begin{equation} t\approx {r\over
2\alpha_ic[\gamma(r)]^2}+{r\theta^2\over 2c}~.  
\end{equation} 
Photons which are emitted from the shock front at $\theta=0$ reach the
observer at a time 
\begin{equation} 
t=r/2\alpha_i c\gamma^2~. 
\end{equation} 
Neglecting redshift effects,
the differential luminosity seen by the observer at time
$t$ is obtained by 
integrating eq. 2 over $r'\leq r$, $r$ and $\theta$, subject to eq. 3
(thick shell) or eq. 2 (thin shell).  Because
the angular delay dominates eqs. 3 and 4,
the emissivity is weighted in the integration by $ 2\pi \theta d\theta$,
the integrand peaks at $\theta^2=1/(6-i+p)\gamma^2 $.  Substituting that
into eq. 4, we find that for thin adiabatic shells ($i=2$) most of the
high frequency photons which arrive at time $t$ come from a ring around
the l.o.s. whose distance $R$ and Lorentz factor $\gamma(R)$ satisfy eq. 5
with $\alpha_2\approx 3.6$ and $R=0.77R_{max}$, while for thin radiative
shells $(i=1)$ we find $\alpha_1\approx 4.9$ and $R \approx 0.84R_{max}$. 
Very similar results were obtained by Panaitescu and Meszaros (1997) from
exact numerical integrations.  

The relativistic expansion lasts until $\gamma(r)\approx 2$, i.e.,
$t\approx r/8\alpha_i c$.  Since the energy of the swept up material is
$\approx(4\pi/3)r^3n\gamma^im_pc^2$, the explosion energy must satisfy
\begin{equation} 
E\geq 2.7\times 10^{54}
n(\alpha_i/[1+z])^3i^2t_y^3~erg,
\end{equation}
where $n$ is the mean density of the swept up ISM in $cm^{-3}$, $t_y$ is
the observer time in years, and $z$ is the redshift of the host galaxy
where the explosion took place.  (The factor $i^2=4$ for the thick
shell/adiabatic expansion case follows from the assumption that the proton
and electron temperatures are both $\sim \gamma m_pc^2$). 
The shape, angular size $(0.8'')$ and magnitude $(V=25.7\pm 0.15)$ of the
host nebula of GRB 970228 that were measured by HST between Sep. 4.65 and
4.76 UT ($t_y\approx 0.52$) suggest that it is a galaxy with a redshift
$z<1$.  For $z=1$, a standard ISM density $n\sim 1~cm^{-3}$, $i=1$ and
$\alpha_1\approx 4 $ calculated by Panaitescu and Meszaros (1997) for a
thin/radiative shell, eq. 6 yields $E\geq 3\times 10^{54}~erg $. For a
thick/adiabatic shell ($i=2$) and $\alpha_2\approx 2$ eq. 6 yields 
$E\geq 1.5\times 10^{54}~erg$. Such energies, are comparable to the
total energy-release in mergers/AIC of compact stellar objects, which is
usually less than $\sim M_\odot c^2\approx 1.8\times 10^{54}erg$. Such
kinetic energies, however,  are larger by orders of magnitude than the
maximal plausible kinetic energies of spherical explosions  
produced by such events. This is because a large
fraction of the released energy is radiated in gravitational waves, and
neutrino emission is inefficient in driving spherical explosions in
gravitational collapse of compact objects. Typically, in core collapse
supernovae explosions, the kinetic energies of the debris is about $\sim
1\%$ of the total gravitational binding energy release.  NS merger/AIC is
not expected to convert a larger fraction of the gravitational binding
energy release into kinetic energy of a spherical explosion.  First, a
large fraction of the binding energy is radiated away by gravitational
waves emission, which is relatively unimportant in Type II supernova
explosions.  Second, neutrino deposition of energy and momentum in the
ejecta is less efficient in NS mergers, because it lasts only for
milliseconds and because neutrino trapping and gravitational redshift of
neutrino energy are stronger than in core collapse supernovae.  Indeed,
detailed numerical calculations of spherical explosions driven by
neutrinos in NS mergers (e.g., Janka and Ruffert 1996;  Ruffert et al
1997) produce very small explosion energies. Although the numerical
calculations still are far from being full general relativistic three
dimensional calculations, let alone their inability to reproduce
consistently supernova explosions, probably, they do indicate the correct
order of magnitude of the kinetic energy in spherical explosions driven by
NS merger or AIC of white dwarfs and NS. 

The fluence of GRB 970508 was $\geq 10^{52}erg $ in the 0.04-2.0 $MeV$
alone, assuming isotropic emission. If hundred times brighter GRBs, like
GRB 970616, have redshifts similar to that of GRB 970508, their fluences
must be $\sim 10^{54}~erg$ for isotropic emission. It also cannot be supplied
by mergers/AIC of compact stellar objects. 

\subsection{Absence of Simple Scaling} 
Relativistic blast wave models predict
that GRB afterglows are scaled by powers of their basic parameters: total
energy E, initial Lorentz factor $\Gamma$, surrounding gas density $n$,
and distance $D$. However, GRBs 970111, 970228, 970402, 970508, 970616,
970828 and 971214 exhibited unscaled behavior and very different spectral
properties (for the X-ray observations see Costa et al. 1997; Piro et al.,
1997;  Castro-Tirado et al 1977;  Feroci et al 1977; Heise et al. 1997;
Odewahn 1997; Frontera et al 1997; for optical observations see the
compilation in Reichart 1997; Sahu et al 1997b; Pedersen et al 1997; and
Halpern et al. 1997; A. Diercks et al. 1997; for radio observations see
Frail et al 1997b and references therein). For instance, GRB 970111 and
GRB 970828 had $\gamma$-ray fluences $\sim$ 25 times larger than GRB
970228 but their afterglow were not detected in X-rays, in the optical
band and in the radio band (e.g., Groot et al. 1997b; Frontera et al.
1997). The upper bound on the optical peak response of GRB 970828 was
$\sim 10^2,~10^3$ smaller than that of GRB 970228 and GRB 970508,
respectively (Groot et al 1997b).  GRB 970508 was 6 times weaker in
$\gamma$-rays than GRB 970228 (Kouveliotou et al 1997, Hurley et al 1977)
but 6 times brighter in the optical band (see, e.g., Sahu et al 1977b and
references therein).  Such spectral variability is observed in the
afterglows of gamma ray flares from extragalactic relativistic jets of
blazars and also in flares from galactic relativistic jets of microquasars
(galactic superluminal sources) such as GRS 1915+105 (Mirabel and
Rodriguez 1994) and GRO J1655-40 (Tingay et al. 1995). 

\subsection{Firecone Rescue?} 
The radiated energy of GRB 970228 during the afterglow in the 2-10 keV
window alone was about 40\% of the energy in the gamma burst itself in the
40-700 keV band (Costa et al.  1997).  For such a fast cooling,
energy-momentum conservation requires $\gamma\sim r^{-3}$, instead of
$\gamma\sim r^{-3/2}$ for a slow cooling, which was used to derive the
$\sim t^{-3(p-1)/4}$ fading of the X-ray and optical afterglows (Wijers et
al. 1997) of GRBs. Also the relation between observer time, emissiom
radius and Lorentz factor which was used is not correct.  Thus, the only
successful prediction of the afterglow model is also in doubt.  Moreover,
the duration (in months) of the initial power-law fading of the afterglow
(thin radiative shell, $i=1$, $\alpha_1\approx 4)$ which last until
$\gamma(R)\approx 2$ is
\begin{equation}
t_m\approx 1.85 E_{52} ^{1/3}[(1+z)/i^{2/3}\alpha_in^{1/3}]~months,
\end{equation} 
where $E_K=10^{52}E_{52}~erg$ and $n$ in $cm^{-3}$.  This short cooling
time is already in conflict with the observed $\sim t^{-1.1}$ fading of
the afterglow of GRB 970228 over 6 months (Fruchter et al. 1997) if
$[E_{52}/n]^{-1/3}\leq 1$, both for thin/radiative and thick/adiabatic
shells. Note that GRBs 970228 and 970508 appear within the optical
luminous part of the faint host galaxy (Sahu et al. 1997, Fruchter 1997,
Metzger et al 1997, Djorgovski et al 1997)  where one expects $n\sim 1$.
Conical fireballs (``firecones'')  with opening angles $\theta_c>1/\Gamma$
and solid angles smaller by $\theta_c^2/4$, can reduce the estimated total
energy in $\gamma$-rays and X-rays by a factor $\sim \theta_c^2/4$.  As
long as $\theta_c>1/\gamma$, fireballs and firecones look alike for
observer near the axis of the firecone. But, when $\gamma\theta_c<1$, the
beaming efficiency decreases by $\gamma^2\theta_c^2$ and the $\sim
t^{-1.1}$ fading of the optical afterglow is accelerated by a factor
$\gamma^2\sim t^{-6/(6+i)}=t^{-3/4}$, for thick/adiabatic conical shell. 
Such a change has not been observed yet in the afterglow of GRB 970228,
implying that after six months $\gamma\theta_c>2$. Therefor, firecones
cannot solve (by additional factors $<4,~4^{1/3}$ on the r.h.s. of eqs. 
6 and 7 , respectively) the energy crisis or explain the uniform power law
fading of GRB 970228 for over six months. It can be shown easily that the
crisis is larger for observers with larger viewing angles with respect to
the firecone axis.

\section{Short Time Variability} 
Even if the energy crisis in GRBs and the non-universality of their
afterglows could have been avoided by assuming firecones, i.e. conical
shells instead of relativisticly expanding spherical shells, neither
firecones nor fireballs can explain subsecond variability in GRBs that
last for tens or hundreds of seconds.  First, a variable central engine
must be fine tuned to arrange for shells to collide only after a distance
where the produced $\gamma$-rays are not reabsorbed, which is larger by
more than 10 orders of magnitude than the size of the central engine
(Shaviv, 1996).  Second, even with fine tuning of the central engine, the
transverse size of the emitting area whose radiation is beamed towards the
observer, $r\theta \leq r/\Gamma$ where $\Gamma\approx \gamma(0)$, implies
variability on time scales (e.g., Shaviv 1996; Fenimore 1996)
\begin{equation} 
\Delta t\sim r\theta^2/2c\approx r/2c\Gamma^2\sim T_{GRB}, 
\end{equation} 
i.e., comparable to the total duration of the GRB. It is in conflict
with the observed short time variability of GRBs.  Even GRBs that last
more than $100~ s$, show a variability on subsecond time scales, 
(e.g., Fishman and Meegan 1995). Local instabilities are not efficient
enough in producing high intensity pulses.

\subsection{Extended GeV Emission} 
The initial Lorentz factor of a relativisticly expanding fireball, which
sweeps up ambient matter, decays rather fastly ($t\sim T_{GRB}$) as its
energy is shared by the swept up matter. It cannot explain emission of
multi GeV $\gamma$-rays, which is extended over hours (in the observer
frame) with an energy fluence similar to that in the keV/MeV GRB, as
observed in GRB 910503 (Dingus et al. 1994) and in GRB 940217 (Hurley et
al. 1994).  Note in particular that inverse Compton scattering of GRB
photons or external photons by the decelerating debris is not efficient
enough in producing the observed extended emission of GeV photons. Also it
cannot explain MeV $\gamma$-ray emission that extends over 2 days, which,
perhaps, was the case if the cluster of four GRBs (Meegan et al. 1996;
Connaughton et al. 1997) were a single GRB. 

\section{GRBS FROM ACCRETION JETS} 

Highly collimated relativistic jets seem to be emitted by all
astrophysical systems where mass is accreted at a high rate from a disk
onto a central (rotating?) black hole. They are observed in galactic and
extragalactic superluminal radio sources, like the galactic microquasars
GRS 1915+105 (Mirabel and Rodriguez 1994) and GRO J1655-40 (Tingay et al.
1995) and in many extragalactic blazars where mass is accreted onto,
respectively, stellar and supermassive rotating black holes. They produce
$\gamma$-ray flares with afterglows in the X-ray, optical and radio bands
which rise fastly and decline with time like a power-law and have a
non-thermal power-law spectra and hardness which is correlated with
intensity. Highly relativistic jets probably are ejected also in the
violent merger/AIC death of close binary systems containing compact
stellar objects.  Such jets which are pointing in our direction can
produce the cosmological GRBs and their afterglows (Dar 1997b,c).  Jetting
the ejecta in merger/AIC of compact stellar objects can solve the energy
crisis of GRBs by reducing the total inferred energy release in GRBs by
the beaming factor $f=\Delta\Omega/4\pi$, where $\Delta\Omega$ is the
solid angle into which the emission is beamed.  In fact, in order to match
the observed GRB rate (e.g. Fishman and Meegan 1995) and the currently
best estimates of the NS-NS merger rate in the Universe (e.g. Lipunov et
al. 1997) solid angles $\Delta\Omega\sim 10^{-2}$ are required. Such solid
angles are typical of superluminal jets from Blazars.  The estimated rate
of AIC of white dwarfs and neutron stars is much higher, $\sim 1$ per
second in the Universe compared with $\sim 1$ per minute for NS-NS
mergers. If GRBs are produced by accretion induced collapse of white
dwarfs and neutron stars (e.g., Goodman et al 1987; Dar et al 1992), then
$\Delta\Omega\sim 10^{-4}$. 
  
The FeII and MgII absorption lines and OII emission lines at redshift
z=0.835 in the afterglow of GRB 970508 seems to indicate that GRBs are
produced in dense stellar regions, e.g. star burst regions. Boosting of
stellar light by superluminal jets from merger/AIC in dense stellar
regions (with typical size $R\approx 10^{18}\times R_{18}~cm$ and photon
column density $N_\gamma=N_{23}\times 10^{23}~cm^{-2}$) has been proposed
by Shaviv and Dar (1995; 1997) as the origin of GRBs.  It can explain
quite naturally the fluence, typical energy, duration distribution, light
curves, spectral evolution and afterglows of GRBs. Due to space
limitation, here we only demonstrate that it solves the main difficulties
of the fireball/firecone models: If the ejected jet (blobs) has an initial
kinetic energy $E_k=E_{52}\times 10^{52}~erg$, a Lorentz factor
$\Gamma=\Gamma_3\times 10^3$, and a cross section $S_j\approx\pi R_j^2=\pi
R_{j16}^2\times 10^{32}~cm^2 $ which after initial expansion remains
constant due to magnetic confinement, then: 

\noindent (a) The photon fluence at a distance $D=D_{28}\times10^{28}~cm$ 
due to photo absorption/emission by partially ionized heavy 
atoms (Shaviv 1996) in the jet ($\sigma_a=\sigma_{18}\times 
10^{-18}~cm^2$) is \begin{equation}  
I_\gamma\approx {\eta E_k\sigma_{T} N_\gamma\over \Gamma m_p c^2 
D^2\Delta\Omega} =7\times {\eta_2E_{52}\sigma_{18}N_{23}\over
                   D_{28}^2\Gamma_3\Delta\Omega_2}~\gamma~cm^{-2},
\end{equation} 
where $\eta=\eta_2\times 10^{-2}$ is the fraction of heavy atoms in 
the jet (we assume a cosmic ray composition). 

\noindent (b) The typical 
energy of the emitted (Lorentz boosted) photons and the energy fluence
in the observer frame are, respectively,  
\begin{equation} 
E_\gamma\approx {\Gamma_3^2\epsilon_{eV}\over(1+z)}~MeV,
\end{equation} 
where $\epsilon=\epsilon_{eV}\times eV$ is the typical energy of stellar
photons, and  
\begin{equation} 
F_\gamma\approx I_\gamma E_\gamma\approx  10^{-5}\times 
{\eta_2E_{52}\sigma_{18}N_{23}\Gamma_3\epsilon_{eV} \over (1+z)
                   D_{28}^2\Delta\Omega_2}~ erg~cm^{-2}.
\end{equation} 
(c) The typical duration of GRBs and the duration of individual pulses from 
boosting starlight of bright stars are given, respectively, by
\begin{equation} 
T_{GRB}\approx {R\over c\Gamma^2}=30 R_{18}\Gamma_3^{-2}~s,~and~~ 
T_p\approx {R_j\over c\Gamma^2}=0.30 R_{j16}\Gamma_3^{-2}~s.
\end{equation} 
The bimodality of the duration distribution og GRBs (e.g., Fishman and 
Meegan 1995) has a simple statistical origin (Shaviv and Dar 1997). 

\noindent(d) Due to energy-momentum conservation, an ejected jet (blob)
with a an initial kinetic energy $E_k$, bulk-motion Lorentz factor
$\Gamma$ and constant cross section $S_j$ is decelerated by the swept up
interstellar matter according to $\gamma=\Gamma/(1+R/R_0)$, or
$R=R_0(\Gamma/\gamma-1)$, where $R$ is its
propagation distance in the interstellar medium and
$R_0=E_k/nm_pc^2\Gamma S_j$. The electrons in the ejecta and the
swept up interstellar matter whose total mass increases like $M\sim
1/\gamma$ are accelerated by the jet to a power-law spectrum in the jet
rest frame. They emit 
synchrotron radiation  with a power-law spectrum $\nu'dn/d\nu'
\sim \nu'^{-(p-1)/2}$ with intensity proportional to their number and
to the magnetic energy density. For an observer within the beaming cone,
this synchrotron emission is Lorentz boosted and collimated
according to eq. 2, i.e., it is amplified by a factor $A\sim 
\gamma^{3+(p-1)/2}$. Thus, an observer within the beaming cone sees
a synchrotron afterglow with intensity $I_{\nu}\sim AB^2M(dt'/dt)$. 
Since $dt=dt'/2\gamma$
and $t=\int 
dR/2\gamma^{-2}=(R_0/6c\Gamma^2)[(\Gamma/\gamma)^{-3}-1]$, one obtains 
for $p=2.5\pm 0.5$ that
 \begin{equation}
I_\nu\sim \gamma^{3+(p-1)/2} \sim (t+t_0)^{-1.25\pm 0.08}
 \end{equation}
where $t_0=R_0/6c\Gamma^2\approx (E_{52}/n\Gamma_3^3R_{j16}^2)\times 
100~s$. 
Initial expansion of the ejecta, changes in opacity within the 
jet and along the trajectory of the emitted radiation,  
and viewing angle effects due to the change in the beaming
angle, can produce complex time and wavelength dependences of 
the afterglow in the initial phase. Moreover, absorption
of optical photons, UV photons and X-rays by the interstellar gas 
and dust around the burst location depends strongly on energy.
Gas Column densities $N_{H}\geq 10^{22}~cm^{-2}$, which are also required 
by the detection of GeV emission from bright GRBs (see below),
can explain why some GRBs afterglows which were detected in $X$ rays 
were not detected also in the optical band. If this explanation for the 
suppression  of optical afterglows of GRBs is correct, then  X-ray 
afterglows of GRBs which are not accompanied by optical afterglows must 
show harder X-ray spectra than those of GRBs with optical afterglows.
Such GRBs must also be accompanied by strong emission of GeV photons.

Inhomogeneous ISM and jet instabilities can modify the late time behavior
of the afterglows. For instance, if the jet is deflected by a stellar or
interstellar magnetic field, the afterglow may disappear suddenly from the
field of view (collisions and deflection of jets on scales 10-100 pc were
observed in AGN, e.g., Mantovani et al. 1997).

\noindent (e) The high column density of gas in star forming regions,
$N_{H}=N_{23}\times 10^{23}~cm^{-2}$, with $N_{23}\geq 1$, provides an
efficient target for hadronic production of high energy photons via
$pp\rightarrow \pi^0 X$ followed by the prompt $\pi^0\rightarrow 2\gamma$
decay. A power-law proton spectrum produces a power-law photon spectrum
with the same power index and efficiency (e.g., Dar 1997)
$g\sigma_{in}N_{H}$ where $g=10^{-1}\times g_1
=0.195exp[-3.84(p-2)+1.220(p-2)^2]$ and $\sigma_{in}\approx 3\times
10^{-26}~cm^2$ is the $pp$ inelastic cross section.  Consequently, GRBs in
star forming regions are accompanied by emission of a power-law spectrum
of high energy photons with a total fluence
\begin{equation}
F(>100~MeV) \approx {E_{52}g_1N_{23}\over D_{28}^2\Delta\Omega_2}{3\over 1+z}
\times 10^{-6} erg~cm^{-2},
\end{equation} 
comparable to the GRB fluence in MeV $\gamma$ rays.
This is consistent with the detection of GeV photons by the EGRET 
instrument on board the Compton Gamma Ray Observatory from a handful of
bright bursts (see, e.g., Dingus 1995). Given the EGRET sensitivity
and limited field of view, the detection rate implies that high energy 
emission may accompany most GRBs. 

Finally, significant hadronic production of gamma rays with energy $\sim
18~ GeV$, as observed in GRB 940217, requires incident proton energies
$\sim 6$ times larger, i.e., that $\gamma>115$. Consequently, the effective
duration of emission of such photons is 
\begin{equation}
t(E_\gamma<18 GeV)\approx {R_0\over 6c\gamma^2}\approx
{E_{52}\over n\Gamma_3R_{j16}^2}\times 2.5h, 
\end{equation}
which is consistent with the EGRET/CGRO observations (Hurley 1994).    

\noindent
\section{CONCLUSIONS} 

The observed properties of GRBs and their afterglows, in particular that
of GRB 970228 six months later, seem to rule out relativistic fireballs
and firecones powered by mergers/AIC of compact stellar objects within
galaxies as the origin of GRBs. In spite of their flexibility and
multitude of free parameters, the simple fireball and firecone models of
GRBs appear not to be able to explain the total energy of GRBs, nor to
explain the enormous diversity of GRBs, their short time scale (subsecond)
variability, their spectral evolution, the delayed emission of MeV and GeV
$\gamma$-rays in some GRBs, and the spectral versatility of GRB
afterglows. In order to solve these problems, the single relativistic
spherical shell which expands into a uniform medium must be replaced by a
fine tuned series of asymmetric shells (conical ejecta) which expand into
a nonuniform medium (e.g., Meszaros et al 1997). This adds many new
parameters to the ``fireball'' model which can be adjusted to fit any GRB
and rescue the model. However, this increased flexibility through a
multitude of new adjustable parameters makes the modified relativistic
fireball/firecone models too flexible, without predictive power and
unfalsifyable, and therefor scientifically unacceptable.  However, if the
relativistic ejecta in merger/AIC of compact stellar objects is collimated
into highly relativistic jets, most of the problems of the spherical
fireball models can be avoided and the general properties of GRBs and
their afterglows can be explained quite naturally using the observed
properties of superluminal jets from blazars and microquasars. In
particular, if GRBs are produced by highly relativistic jets which are
pointing in our direction they should show superluminal motions with
speeds $v\leq \Gamma c$.  Such supeluminal motions may be detected in long
term (months) VLBI observations of radio afterglows of GRBs (see, e.g.,
Taylor et al. 1997).

\end{document}